\documentclass[preprint,...]{revtex4-1}

\usepackage{graphicx}
\usepackage{dcolumn}
\usepackage{bm}
\usepackage{amsfonts}

\begin{document}


\title{Towards the efficiency limits of silicon solar cells: how thin is too thin?}

\author{Piotr Kowalczewski}
\email{piotr.kowalczewski@unipv.it}
\author{Lucio Claudio Andreani}
\affiliation{Department of Physics and CNISM, University of Pavia, Via Bassi 6, I-27100 Pavia, Italy}

\date{\today}

\begin{abstract}
It is currently possible to fabricate crystalline silicon solar cells with the absorber thickness ranging from a few hundreds of micrometers (conventional wafer-based cells) to devices as thin as $1\,\mu\mathrm{m}$. In this work, we use a model single-junction solar cell to calculate the limits of energy conversion efficiency and estimate the optimal absorber thickness. The limiting efficiency for cells in the thickness range between 40 and  $500\,\mu\mathrm{m}$ is very similar and close to 29\%. In this regard, we argue that decreasing the thickness below around $40\,\mu\mathrm{m}$ is counter-productive, as it significantly reduces the maximum achievable efficiency, even when optimal light trapping is implemented. We analyse the roles of incomplete light trapping and extrinsic (bulk and surface) recombination mechanisms. For a reasonably high material quality, consistent with present-day fabrication techniques, the optimal thickness is always higher than a few tens of micrometers. We identify incomplete light trapping and parasitic losses as a major roadblock in improving the efficiency upon the current record of 25.6\% for silicon solar cells. Finally, considering the main parameters that impact solar cell performance, we quantify the constraints and requirements for achieving a specified energy conversion efficiency, which is important for a proper design strategy of high efficiency silicon solar cells.

\end{abstract}

\pacs{Valid PACS appear here}
\maketitle

\section{Introduction}
\label{sec:intro}

In this work we focus on the efficiency limits of crystalline silicon (c-Si) solar cells. In this regard, we consider both ideal devices (perfect material and interfaces), as well as more realistic conditions, including defect-related recombinations, parasitic losses, and non-optimal light management. It is currently possible to fabricate c-Si solar cells with absorber thickness values that differ by two orders of magnitude. On the one end of the thickness range there are conventional wafer-based c-Si solar cells with the absorber thickness of the order of a few hundreds of micrometers \cite{Zhao1998}. Epitaxial growth allows fabricating solar cells with the thickness of a few tens of micrometers \cite{Petermann2012}. Finally, epitaxy-free fabrication\cite{Meng2012a,Trompoukis2012} makes possible fabricating c-Si solar cells with the absorbing layer as thin as $1\,\mu\mathrm{m}$. For a proper design strategy, it is important to determine the optimal absorber thickness in terms of the energy conversion efficiency. We address this question using a model single-junction silicon solar cell.

It is well know that reducing the silicon thickness leads to the reduction of total absorption, and thus of the photocurrent. This has to be compensated by implementing an appropriate light-trapping scheme. Yet, it should be emphasized that even when the optimal light trapping is applied (i.e., perfect anti-reflection action combined with a Lambertian scatterer), the maximum achievable absorption still decreases with decreasing material thickness \cite{bozzola2012photonic,bozzola2014towards}. On the other hand, $V_\mathrm{oc}$ generally tends to decrease with increasing thickness. For a given material quality, this leads to an optimal thickness that maximizes the conversion efficiency \cite{bozzola2014towards,kowalczewski2014light}. In this context, the goal of this work is twofold: First, to establish the most suitable thickness range for c-Si solar cells to approach the efficiency limits. Second, to identify the parameters that have to be improved in order to increase the energy conversion efficiency beyond the current record of 25.6\% \cite{hit2014}.

We use efficiency as a figure of merit to assess different solar cell structures. This is motivated by the fact that the cost of electricity is mainly determined by the efficiency rather than by the cost of the active material (which, in the case of silicon, is constantly decreasing). In this regard, it is particularly important to estimate the optimal absorber thickness range that maximizes the efficiency. In our analysis we include intrinsic Auger recombination, as well as defect-based bulk and surface recombination mechanisms. We also introduce a simple approach that allows us to consider parasitic losses. We compare our results with the measured performance of state-of-the-art silicon solar cells, showing the room for improvement in terms of current, voltage, and fill factor. We pay particular attention to parasitic losses, which we believe are a major roadblock in reaching efficiency above the current record of 25.6\% \cite{hit2014}.

A review of the efficiency limits of silicon solar cells is given in Ref.~\cite{swanson2005approaching}. The limits reported in the literature \cite{Tiedje1984,kerr2003limiting,richter2013reassessment} are usually calculated using the idealized diode equation \cite{Nelson2003}. This approach has the following limitations:

\begin{enumerate}
\item The diode equation gives less accurate results when the cell thickness is decreasing. We attribute this inaccuracy to the assumptions underlying the treatment of the space-charge region (SCR) in the idealized diode formalism. In practice, idealized diode equation tends to significantly overestimate efficiency for thin cells. We further discuss the accuracy of the results obtained using the idealized diode equation in the Appendix.
\item Solar cells require selective contacts, which can be achieved using a $p$-$n$ junction. Yet, the junction is not explicitly considered in the ideal diode approach. Therefore, ideal diode equation gives unrealistic results that overestimate efficiency for undoped silicon.
\end{enumerate}

\noindent In our approach, we calculate the efficiency limits overcoming these limitations: we analytically obtain the photogeneration rate (ranging from double-pass absorption to Lambertian light trapping), explicitly consider a $p$-$n$ junction, and numerically solve the drift-diffusion equations. This allows us to calculate more realistic efficiency limits of silicon solar cells in a wide range of cell thicknesses and doping levels. 

The paper is organized as follows: in Sec.~\ref{sec:num_approach} we describe our approach, based on analytical photogeneration rate and numerical solution of drift-diffusion equations. In Sec.~\ref{sec:limits} we calculate the efficiency limits for c-Si solar cells. In Sec.~\ref{sec:lt} we discuss the effects of incomplete light trapping in approaching the efficiency limits. In Sec.~\ref{sec:srh} we cover the impact of bulk material imperfections on the cell performance, and introduce a simple approach to include parasitics losses in the analysis. In Sec.~\ref{sec:sr} we discuss the role of surface recombination. In Sec.~\ref{sec:requirements} we quantify the requirements, in terms of bulk and surface material quality, to achieve a given efficiency level. Conclusions are given in Sec.~\ref{sec:conclusions}. Finally, in the Appendix we compare the results of our numerical treatment with those obtained using the idealized diode equation.

\begin{figure}[!t]
\includegraphics[width=0.6\textwidth]{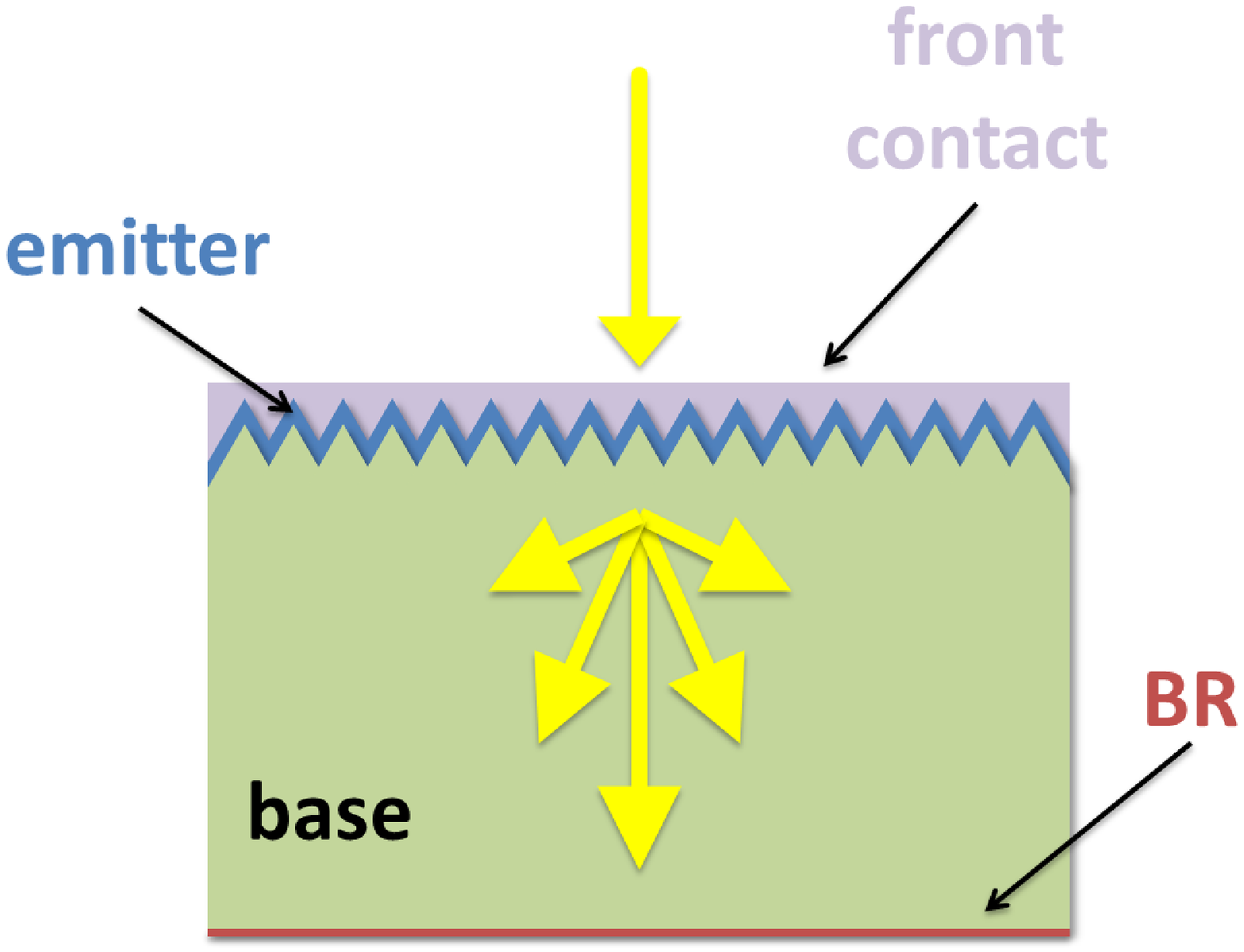}
\caption{\label{fig:fig1} Investigated solar cell structure consisting of a $5\,\mathrm{nm}$ thick $n$-type emitter (the blue region), $p$-type base (the green region), and a perfect back reflector (BR), which also serves as a back contact. The front surface is textured so that the incident light is scattered and trapped within the absorber.}
\end{figure}

\section{Numerical approach}
\label{sec:num_approach}

Let us consider the structure sketched in Fig.~\ref{fig:fig1}. It consists of a $5\,\mathrm{nm}$ thick $n$-type emitter and a $p$-type base of variable thickness. Such a thin emitter allows to minimise recombination losses in this heavily doped layer. To calculate the efficiency limits we assume no reflection at the front interface, a perfect back reflector (BR), and a Lambertian light trapping \cite{Yablonovitch1982,Green2002}. In Fig.~\ref{fig:fig1} we schematically show the Lambertian scatterer at the front, yet we note that the photogeneration rate is calculated analytically and in the electrical calculations the structure is assumed to be flat. Finally, we assume full-area contacts: the carriers are collected at the silicon/BR and emitter/front contact interfaces.

The photogeneration rate corresponding to the Lambertian limit is calculated as in Ref.~\cite{bozzola2014towards}:

\begin{equation}
G_\mathrm{LL}(z,E) = \frac{\alpha_\mathrm{lt}\left( R_\mathrm{b} e^{-2\alpha_\mathrm{lt}w} e^{\alpha_\mathrm{lt}z}+e^{-\alpha_\mathrm{lt}z} \right)}{1-e^{-2\alpha_\mathrm{lt}w} \left( 1 - \frac{1}{n_\mathrm{Si}^2}\right)} \times \phi_\mathrm{AM1.5G},
\label{eq:G_LL}
\end{equation}

\noindent where $n_\mathrm{Si}$ is the refractive index of c-Si \cite{green2008self} and $\phi_\mathrm{AM1.5G}$ is the photon flux density corresponding to the AM1.5G solar spectrum \cite{AM15G}. The effective absorption coefficient $\alpha_\mathrm{lt}$ is related to the light path enhancement in textured cells according to Ref.~\cite{Green2002}. In the literature we can find a number of different strategies that allow approaching the Lambertian limit, including ordered \cite{bozzola2012photonic,abass2012dual,Isabella2013,gomard2013photonic} and quasi-ordered \cite{peretti2013absorption,Martins2013Deterministic,Bozzola2013} photonic structures, as well as random textures \cite{Jager2013,battaglia2012lambertian,wiesendanger2013path,Kowalczewski2013}.

The photogeneration rate calculated using Eq.~\ref{eq:G_LL} is integrated with respect to energy (over the solar spectrum) and used as the generation term in the drift-diffusion equations, which are solved using Finite-Element Method (FEM). We use FEM implementation in the commercial device simulator Silvaco Atlas \cite{Silvaco}. This methodology is similar to the one described in our previous works \cite{kowalczewski2014light,andreani2014photonic}.

\section{Efficiency limits}
\label{sec:limits}

We start by considering solar cells limited by intrinsic Auger recombination. We treat Auger recombination using the parametrization reported in Ref.~\cite{richter2012improved}. We also consider band gap narrowing (BGN) according to the model by Schenk \cite{schenk1998finite}. Finally, we neglect free carrier absorption, which is a second-order effect \cite{Tiedje1984,richter2013reassessment}. Silicon is an indirect band gap material, and thus we also neglect losses related to radiative recombination, which may give an appreciable effect only for very thick cells. Yet, for the thick cells the probability of photon recycling increases \cite{kerr2003limiting}, i.e., radiatively emitted photons are reabsorbed. These two effects are likely to compensate each other. In the appendix we show that radiative recombination together with photon recycling have a negligible effect on the cell performance.

\begin{figure}[!p]
\includegraphics[width=0.55\textwidth]{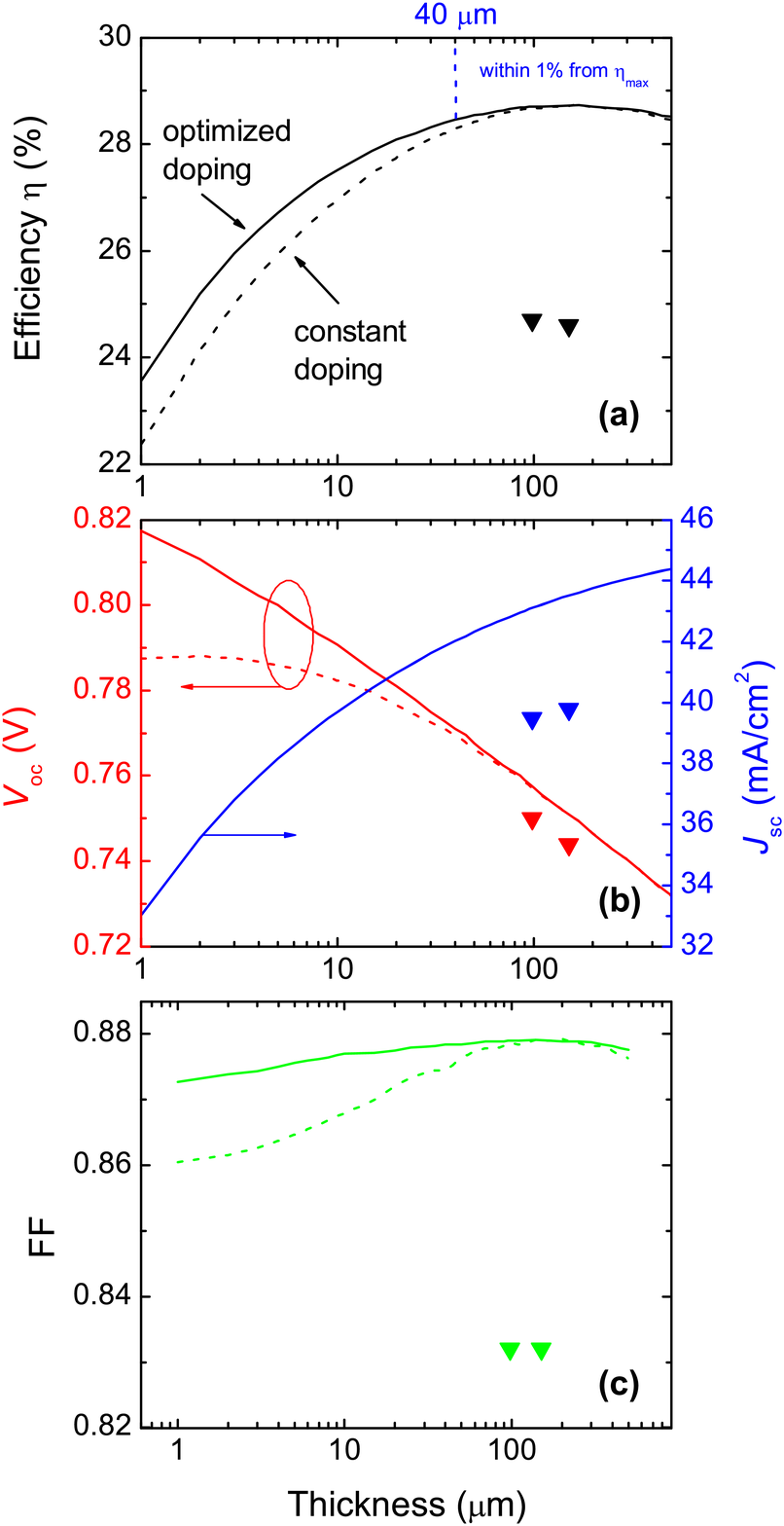}
\caption{\label{fig:fig2} (a) The limiting efficiency of c-Si solar cells as a function of the absorber thickness. (b) $V_\mathrm{oc}$ and $J_\mathrm{sc}$, and (c) fill factor (FF) corresponding to the calculated efficiency limits. The solid lines are calculated by optimizing the base doping for each thickness, whereas the dashed lines are calculated assuming the constant doping $N_\mathrm{a}=10^{16}\,\mathrm{cm}^{-3}$. The triangles denote the performance of the record-efficiency HIT cells \cite{Taguchi2014_24.7}.}
\end{figure}

In Fig.~\ref{fig:fig2}(a) we show the limiting efficiency of c-Si solar cells as a function of the absorber thickness, calculated for the structure sketched in Fig.~\ref{fig:fig1}. For each thickness, we have simultaneously optimized the emitter and base doping concentrations (the doping profile in each layer is assumed to be constant). The efficiency as a function of emitter doping $N_\mathrm{d}$ has a wide maximum, and the optimal value $N_\mathrm{d}=1.5\times10^{18}\,\mathrm{cm}^{-3}$ does not change with the absorber thickness, as the emitter thickness itself is kept constant. Regarding the base doping $N_\mathrm{a}$, the solid lines in Fig.~\ref{fig:fig2} are calculated by optimizing $N_\mathrm{a}$ for each thickness. On the other hand, the dashed lines are calculated assuming a constant doping $N_\mathrm{a}=10^{16}\,\mathrm{cm}^{-3}$, which is the optimal doping for the optimal absorber thickness equal to $170\,\mu\mathrm{m}$. The optimal base doping is highest for thinner cells (reaching $10^{17}\,\mathrm{cm}^{-3}$ for the $1\,\mu\mathrm{m}$ thick cell) and decreases with increasing thickness.

The maximum efficiency is equal to $\eta_\mathrm{max} = 28.73\%$, and it is obtained for the $170\,\mu\mathrm{m}$ thick cell. This is lower than the efficiency limit of 29.43\% reported recently in the literature \cite{richter2013reassessment}. The main reason is that in our calculations, the base doping increases Auger losses and therefore reduces the efficiency. In the contrary, the limit reported in the literature has been obtained using the idealized diode equation and assuming undoped silicon, as explained in Sec.~\ref{sec:intro} and in the Appendix.

We note that the efficiency as a function of the absorber thickness has a broad maximum, and even a small change in the input parameters can substantially shift the nominal optimal thickness. In the range between 40 and $500\,\mu\mathrm{m}$, the calculated efficiency differs by no more than $1\%$ (relative units) from $\eta_\mathrm{max}$. This somehow arbitrary interval shows that for thickness values that differ considerably (by one order of magnitude), the maximum achievable efficiency is very similar. Finally, for an absorber thickness below $40\,\mu\mathrm{m}$, the efficiency drops significantly.

These results are compared with the efficiencies of the state-of-the-art HIT cells \cite{Taguchi2014_24.7} (the triangles). We note that the efficiency of the HIT solar cell has been recently increased to $25.6\,\%$ \cite{hit2014}. Yet, the thickness of the record cell is not specified, and therefore we were unable to include it in Fig.~\ref{fig:fig2}. This analysis shows that the theoretical margin for improvement of single-junction silicon solar cells is around 3\% (absolute units).

\begin{figure}[!t]
\includegraphics[width=0.6\textwidth]{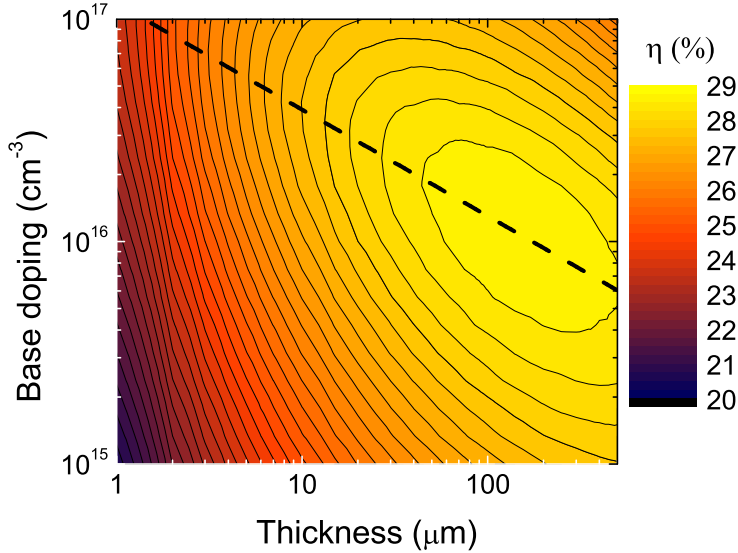}
\caption{\label{fig:fig3} Efficiency as a function of the base doping and absorber thickness. The dashed line indicates the optimal doping.}
\end{figure}

In Fig.~\ref{fig:fig2}(b) we show $V_\mathrm{oc}$ and $J_\mathrm{sc}$ corresponding to the calculated efficiency limits. The base doping optimization allows to slightly improve $V_\mathrm{oc}$, whereas $J_\mathrm{sc}$ is practically the same. Comparing these results with the performance of the HIT cells, we can see that for both values of thickness (98 and $151\,\mu\mathrm{m}$), the measured value is around 99\% of the limiting $V_\mathrm{oc}$, around 94.6\% of the limiting FF, and around 91.6\% of the limiting $J_\mathrm{sc}$. Therefore, these cells are limited by, in decreasing order of significance, $J_\mathrm{sc}$, FF, and $V_\mathrm{oc}$. This shows that optimizing light trapping and, especially, minimizing optical parasitic losses (affecting $J_\mathrm{sc}$), as well as improving the contacts (affecting FF) is the key to further increase the efficiency towards the limiting values.

To further elaborate the issue of the base doping optimization, in Fig.~\ref{fig:fig3} we show the efficiency as a function of the base doping and absorber thickness. This plot can be compared with the results presented in Fig.~4 reported in Ref.~\cite{richter2013reassessment}, which are obtained using the idealized diode equation. In the case of the diode equation, solar cells approach the limiting efficiency for a practically undoped silicon. Yet, in our calculations we can see a clear maximum around $10^{16}\,\mathrm{cm}^{-3}$ (for the thickness range around 100 -- $200\,\mu\mathrm{m}$). Then, efficiency  decreases with decreasing doping. This difference is due to the fact that the junction is not explicitly considered in the ideal diode approach. Therefore, the requirement of selective contacts, which can be achieved using a $p$-$n$ junction, is not included. This leads to unrealistic results for very lightly doped materials.

For the simplicity of the analysis, in the rest of this paper we assume the constant base doping $N_\mathrm{a}=10^{16}\,\mathrm{cm}^{-3}$. 

\section{Effects of incomplete light trapping}
\label{sec:lt}

In the calculations above we have assumed a Lambertian light trapping, which is often taken as a benchmark in the optical design of solar cells. Yet, it is difficult to fulfil this assumption in realistic devices. Therefore, let us now focus on the role of light trapping in achieving the efficiency limits. To do so, we consider solar cells with incomplete light trapping: the photogeneration rate is taken as a weighted average of the photogeneration $G_\mathrm{2p}$ corresponding to the double-pass absorption (i.e., unstructured cell) and the photogeneration $G_\mathrm{LL}$ corresponding to the Lambertian limit. $G_\mathrm{2p}$ is calculated as

\begin{equation}
G_\mathrm{2p}(z,E) = \alpha \left(			e^{-z\alpha}			+ e^{-(2w-z)\alpha}	\right) \times  \phi_\mathrm{AM1.5G}, 
\label{eq:GG_2p}
\end{equation}


\noindent where $\alpha$ is the absorption coefficient of c-Si \cite{green2008self}. As previously, also in the double-pass case we assume a perfect back reflector and anti-reflection action. We note that there are no interference effects included in Eq.~(\ref{eq:GG_2p}). When we consider single wavelengths, the interference effects are profound. Yet, when we integrate the photogeneration rate over solar spectrum, the interference peaks are smeared out, which justifies the approximation used in Eq.~(\ref{eq:GG_2p}). The resulting weighted  photogeneration rate is calculated as

\begin{equation}
G(z) = P_\mathrm{loss} \times \left[ (1-\mathrm{LF}) \times G_\mathrm{2p}(z) + \mathrm{LF} \times G_\mathrm{LL}(z) \right],
\label{eq:phgen}
\end{equation}

\noindent where $\mathrm{LF}$ is the light-trapping factor: $\mathrm{LF}=0$ corresponds to the double-pass case, whereas $\mathrm{LF}=1$ corresponds to the Lambertian limit. $G$ is a function of depth $z$. Finally, $P_\mathrm{loss}$ is a factor related to reflection at the front interface and to parasitic optical losses. For simplicity, we assume that $P_\mathrm{loss}$ does not depend on wavelength. 

\begin{figure}[!t]
\includegraphics[width=0.6\textwidth]{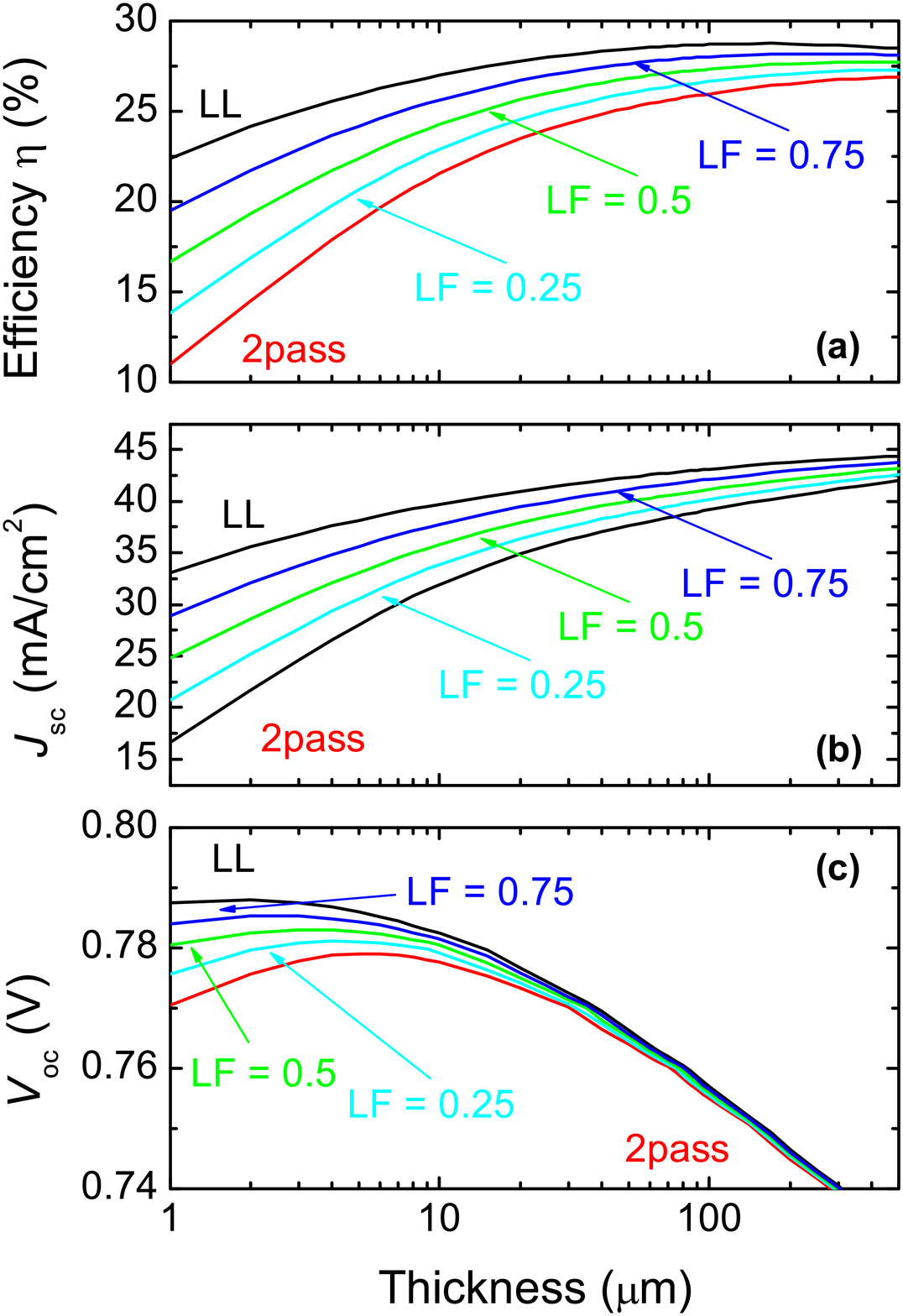}
\caption{\label{fig:fig4} (a) Efficiency, (b) short-circuit current $J_\mathrm{sc}$, (c) open-circuit voltage $V_\mathrm{oc}$ calculated as a function of the absorber thickness for different values of the light-trapping factor $\mathrm{LF}$.}
\end{figure}

We begin by assuming no parasitic losses, that is $P_\mathrm{loss}=1$. In Fig.~\ref{fig:fig4}(a) we demonstrate that light trapping increases the maximum achievable efficiency. This is not the case for semiconductors with a direct energy band-gap, like GaAs, where surface texturing does not increase the maximum efficiency~\cite{miller2012strong}. Therefore, for silicon solar cells light trapping is an essential element required to approach the efficiency limits, even for very thick cells.

As expected, light trapping significantly increases $J_\mathrm{sc}$, which is shown in Fig.~\ref{fig:fig4}(b). $J_\mathrm{sc}$ for the thickest cells does not saturate because of the small absorption below the energy band-gap. Moreover, Fig.~\ref{fig:fig4}(c) shows that light trapping also slightly improves $V_\mathrm{oc}$. Yet, this effect is appreciable only for cell thicknesses below $10\,\mu\mathrm{m}$, that is in a thickness range which is not particularly promising for achieving high efficiency. $V_\mathrm{oc}$ as a function of the thickness for the double-pass case exhibits a gentle maximum. This may be partly because the base doping is not optimized for thin cells.

Although the importance of light trapping for maximizing the conversion efficiency of c-Si solar cells is well known, the present results allow to quantify the effects of incomplete light trapping on the efficiency. Moreover, Eq.~(\ref{eq:phgen}) represents a simple model that can be also used to study the effects of parasitic losses, which are different from the effects related to incomplete light trapping. We shall discuss it in more details in the next section.

\section{Material imperfections and parasitic losses}
\label{sec:srh}

So far, we have considered an idealized material, and therefore the cell performance was limited by intrinsic Auger recombination. Let us now consider extrinsic losses related to defect-base SRH recombination and parasitic optical losses. 

\begin{figure}[!t]
\includegraphics[width=0.7\textwidth]{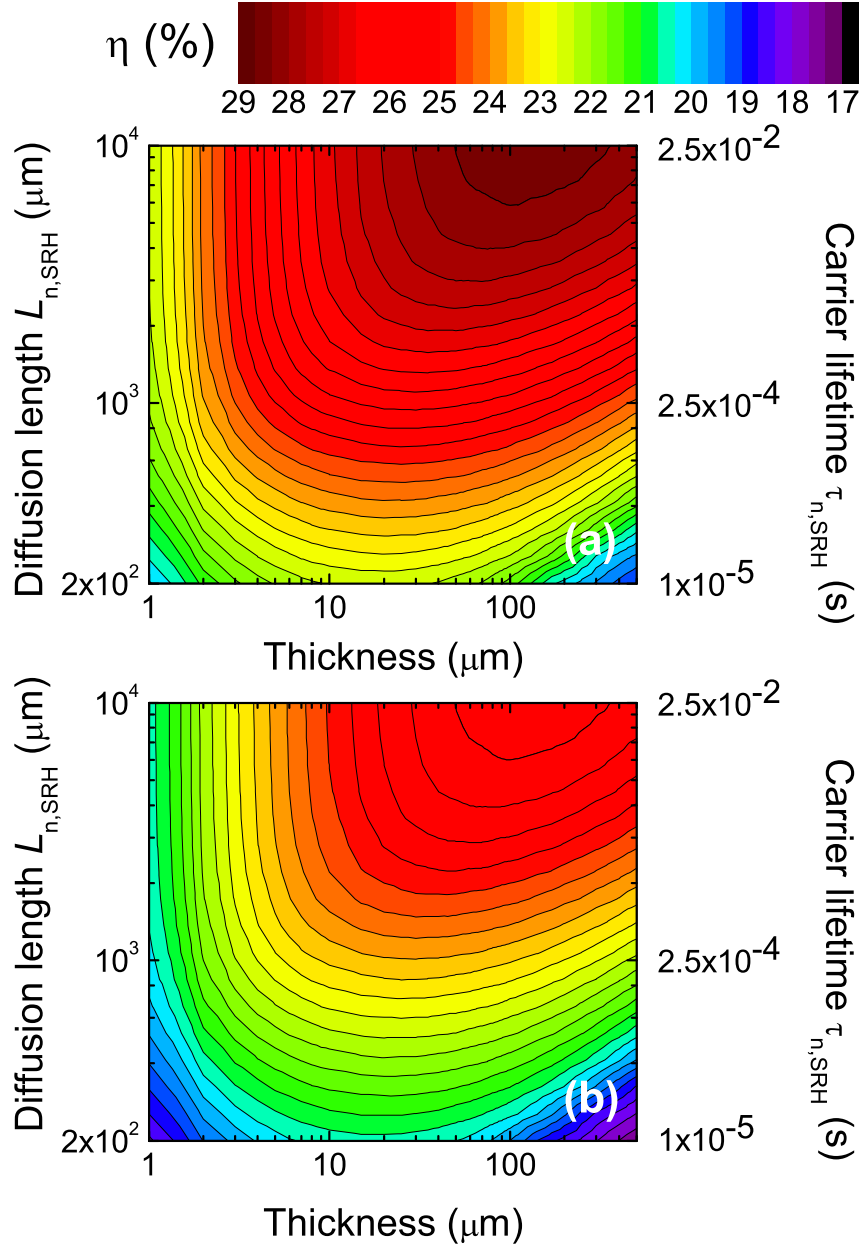}
\caption{\label{fig:fig5} (a) Efficiency as a function of the thickness and diffusion length $L_\mathrm{n,SRH}$ of the minority carriers (electrons) in the base, calculated for the structures with a Lambertian light trapping. (b) The same quantity but including factor related to parasitic losses $P_\mathrm{loss}=0.9$. We note that $L_\mathrm{n,SRH}$ is related to SRH recombination, originating from defects in the base. The scale on the right shows the corresponding carrier lifetime, calculated assuming electron mobility $\mu_\mathrm{n} = 1548\,\mathrm{cm}^2\mathrm{V}^{-1}\mathrm{s}^{-1}$.}
\end{figure}

With the emitter as thin as $5\,\mathrm{nm}$, the cell efficiency is likely to be limited by the diffusion length $L_\mathrm{n,SRH}$ of the minority carriers (electrons) in the base. In Fig.~\ref{fig:fig5}(a) we show the efficiency as a function of the absorber thickness and $L_\mathrm{n,SRH}$, calculated for the structure with Lambertian light trapping. The diffusion length of holes in the highly-doped emitter is taken equal to $L_\mathrm{p,SRH} = L_\mathrm{n}/10$.  It can be seen that thicker cells are more sensitive to SRH recombination. In this regard, thinner cells with the absorber thickness of around $40\,\mu\mathrm{m}$ can reach nearly the same $\eta_\mathrm{max}$ as the thicker cells, but the conditions in terms of absorber quality are relaxed. Finally, reducing the absorber quality (i.e., decreasing $L_\mathrm{n,SRH}$) shifts the optimal thickness towards thinner cells.

Results shown in Fig.~\ref{fig:fig5}(b) are calculated including parasitic losses $P_\mathrm{loss}=0.9$. We estimate that this is approximately the level of parasitic losses in the HIT cells discussed above. The trends are similar to those calculated assuming $P_\mathrm{loss}=1$, yet the maximum achievable efficiency is proportionally decreased. Moreover, for a given efficiency level, the optimal thickness is increased.

The results of Fig.~\ref{fig:fig5} allow us to estimate the optimum thickness and the required material quality to reach a given efficiency level. For example, to surpass 26\% efficiency, in the idealized case of no parasitic losses and complete (Lambertian) light trapping, we need an electron diffusion length around $1\,\mathrm{mm}$. In the more realistic case, when parasitic losses are included (say, $P_\mathrm{loss}=0.9$), achieving efficiency above 26\%  requires a diffusion length higher than $\approx4\,\mathrm{mm}$ for an optimal thickness of $\approx 50\,\mu\mathrm{m}$.

\newpage~

\section{Surface recombination}
\label{sec:sr}

\begin{figure}[!t]
\includegraphics[width=0.7\textwidth]{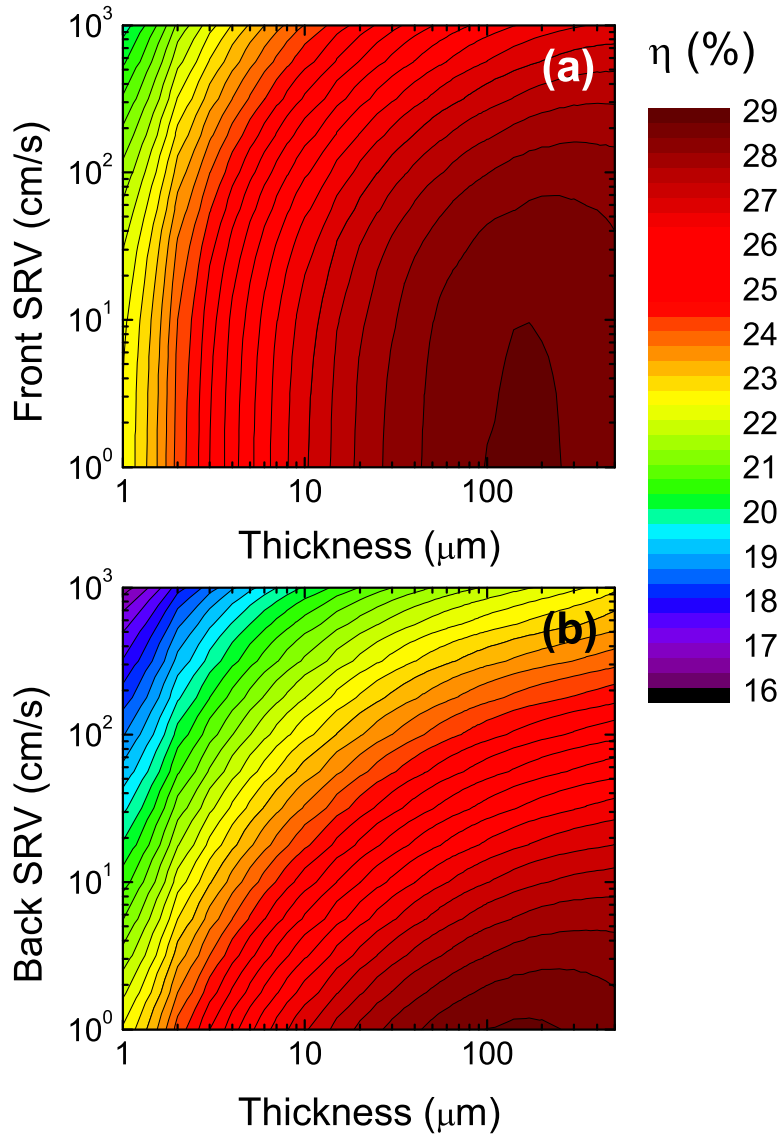}
\caption{\label{fig:fig6} Efficiency as a function of (a) front and (b) back surface recombination velocity (SRV), and of the absorber thickness. For the bulk transport losses, we assume only Auger recombination.}
\end{figure}

\begin{figure}[!t]
\includegraphics[width=0.7\textwidth]{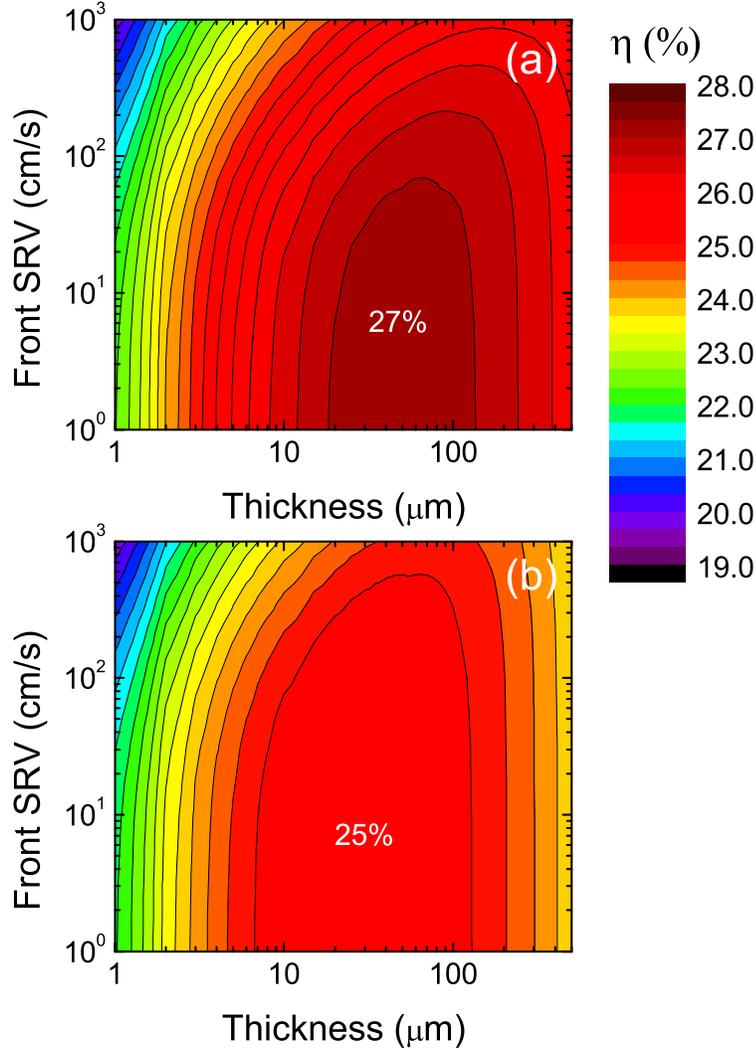}
\caption{\label{fig:fig7} Efficiency as a function of the front surface recombination velocity (SRV) and the absorber thickness. For the bulk transport losses, we assume Auger and SRH recombination. The material quality (i.e., $L_\mathrm{n,SRH}$) is reduced, so that the maximum achievable efficiency is 27\% for $L_\mathrm{n}\approx2200 \,\mu\mathrm{m}$ (a) and 25\% for $L_\mathrm{n}\approx 840\,\mu\mathrm{m}$ (b).}
\end{figure}

Let us now estimate the constraints on the efficiency imposed by surface recombination, which is another extrinsic loss mechanism related to defect states at the surface. In Fig.~\ref{fig:fig6} we show the efficiency as a function of (a) front and (b) back surface recombination velocity (SRV), and of the absorber thickness. At first, for the bulk transport losses we assume only Auger recombination. It can be seen that approaching the efficiency limits requires SRV to be less than a few cm/s, with solar cells being more sensitive to recombination at the rear interface rather than to recombination at the front.

In the calculations above, we assumed a perfect material quality. Yet,  in our previous work we have demonstrated that the importance of surface recombination depends on bulk recombination rate, i.e., the higher is the material quality, the more important are losses at the surface \cite{kowalczewski2014light}. From now on we focus on front surface recombination, neglecting recombination at the back, which has anyway to be of the order of a few cm/s to target high efficiency. In realistic devices, there is usually an oxide passivating layer at the back and point contacts are implemented. For this reason, back SRV can be significantly reduced. On the other hand, front SRV may be increased due to the texturing.

In Fig.~\ref{fig:fig7} we plot the efficiency as a function of the front surface recombination velocity (SRV) and absorber thickness. For bulk transport losses, we assume Auger and SRH recombination. We adjust the diffusion length $L_\mathrm{n,SRH}$ of the minority carriers (electrons) in the base, so that the maximum achievable efficiency is reduced to a certain value. This kind of analysis allows to answer the following question: Having a given material quality, what is the maximum allowed value of the front SRV to achieve the limiting efficiency?

In Fig.~\ref{fig:fig7}(a) the maximum possible efficiency is around 27\%, which is obtained for $L_\mathrm{n}\approx2200 \,\mu\mathrm{m}$. With this material quality, the optimal absorber thickness is $40\,\mu\mathrm{m}$, and the maximum efficiency can be achieved for the thickness range between 20 and  $80\,\mu\mathrm{m}$. In this case, front SRV should be below a few tens of cm/s.

In Fig.~\ref{fig:fig7}(b) we further reduce diffusion length to $L_\mathrm{n}\approx 840\,\mu\mathrm{m}$, so that the maximum possible efficiency is 25\%. The region of highest efficiency (25\%  or slightly above) is clearly wider than in the previous plot. Also the conditions for surface recombination at the front interface are relaxed: in this case, front SRV should be below a few hundreds of cm/s. This analysis confirms that the impact of surface recombination decreases with decreasing material quality. 


\section{Requirements for approaching the efficiency limits}
\label{sec:requirements}

\begin{figure}[!t]
\includegraphics[width=0.7\textwidth]{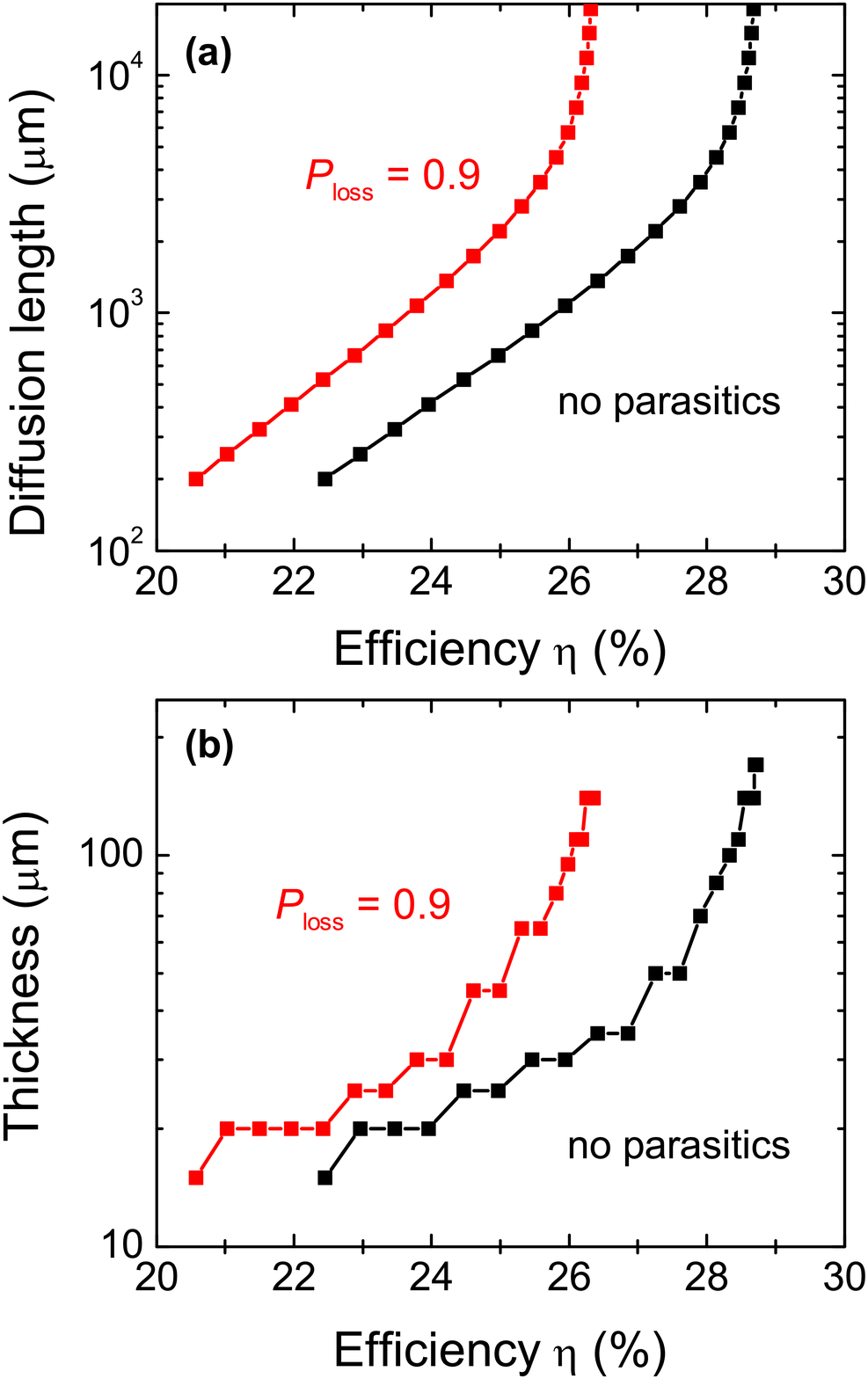}
\caption{\label{fig:fig8} (a) Diffusion length of electrons in the base $L_\mathrm{n}$ as a function of the maximum achievable efficiency. (b) Corresponding (optimal) thickness of the absorbing layer.}
\end{figure}

To summarize the results presented in this paper, in this section we estimate the requirements for approaching a given energy conversion efficiency. In this regard, we analyse both bulk material quality (related to diffusion length of electrons in the base $L_\mathrm{n,SRH}$) and quality of the surfaces (related to surface recombination velocity).

In Fig.~\ref{fig:fig8}(a) we show the diffusion length of electrons in the base $L_\mathrm{n,SRH}$ as a function of the maximum achievable efficiency, whereas Fig.~\ref{fig:fig8}(b) indicates the corresponding (optimal) thickness of the absorbing layer. These results are extracted from Fig.~\ref{fig:fig5}, i.e., for each $L_\mathrm{n,SRH}$ we have extracted the maximum achievable efficiency and the corresponding cell thickness.

In Fig.~\ref{fig:fig8}(a) we can distinguish two trends. In the case of no parasitic losses (black line), for the efficiency range below around 28\%, $L_\mathrm{n}$ and the corresponding maximum achievable efficiency scale linearly. Yet, above 28\%, significant increase of $L_\mathrm{n,SRH}$ gives only a minor, if any, improvement of efficiency.

\begin{figure}[!p]
\includegraphics[width=0.7\textwidth]{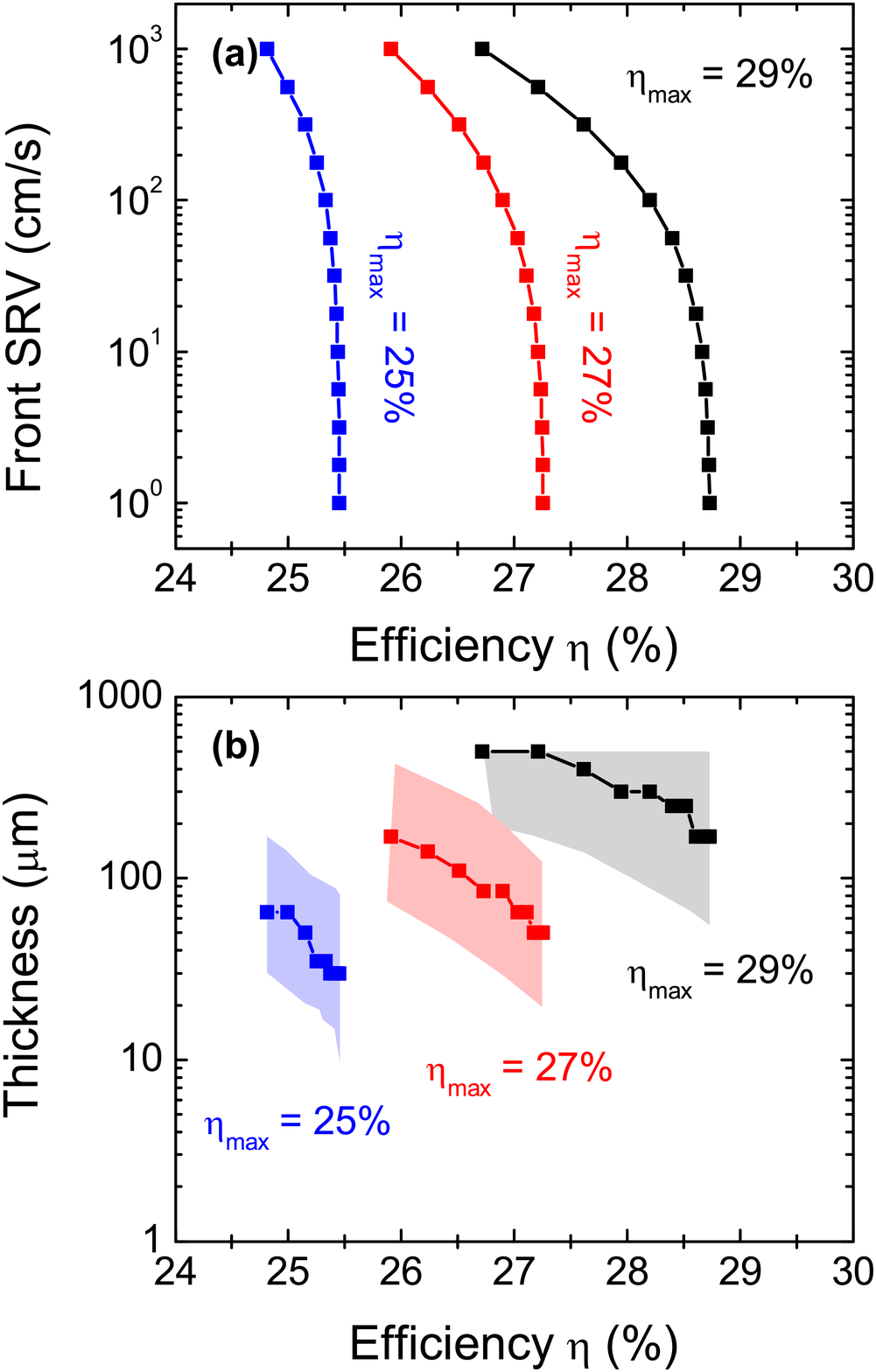}
\caption{\label{fig:fig9} (a) Front surface recombination velocity (SRV) as a function of the maximum achievable efficiency. (b) Corresponding optimal thickness of the absorbing layer. We consider cells limited by Auger recombination (the black lines), as well as cells with SRH recombination: $L_\mathrm{n}\approx1700 \,\mu\mathrm{m}$ with maximum achievable efficiency around 27\% (the red lines) and $L_\mathrm{n}\approx 660\,\mu\mathrm{m}$ with maximum achievable efficiency around 25\% (the blue lines). The colour regions indicate 1\% (relative) tolerance intervals for each material quality.}
\end{figure}

The red line in Fig.~\ref{fig:fig8}(a) shows the results calculated including parasitic losses, that is $P_\mathrm{loss}=0.9$. Parasitic losses result in a shift of the red curve towards lower efficiencies, with the maximum efficiency of around  26\%. Indeed, the measured record efficiency is slightly less than 26\%, which identifies parasitic losses as a major roadblock for going beyond 26\%.

We can also see that an optimal absorber thickness changes significantly in the considered efficiency range, as shown in Fig.~\ref{fig:fig8}(b). For small $L_\mathrm{n}$, the optimal thickness is of the order of a few tens of micrometers. Then, with increasing $L_\mathrm{n}$, the optimal thickness increases rapidly, reaching around $170\,\mu\mathrm{m}$ for $L_\mathrm{n}$ of the order of millimetres.

A similar analysis can be performed regarding surface recombination. In Fig.~\ref{fig:fig9}(a) we show front surface recombination velocity (SRV) as a function of the maximum achievable efficiency, and in Fig.~\ref{fig:fig9}(b) we show the corresponding optimal thickness of the absorbing layer. These results have been extracted from Fig.~\ref{fig:fig6} and \ref{fig:fig7}.

We consider cells limited by Auger recombination (the black lines), as well as cells with SRH recombination: $L_\mathrm{n}\approx2200 \,\mu\mathrm{m}$ with maximum achievable efficiency around 27\% (the red line) and $L_\mathrm{n}\approx 840\,\mu\mathrm{m}$ with maximum achievable efficiency around 25\% (the blue lines). In general, the cells are fairly insensitive to recombination at the front interface if SRV is below a few hundreds of cm/s. This is of the order of magnitude of the effective SRV, currently measured for nanostructured solar cells \cite{Ingenito2015Nanocones}. Moreover, increasing SRV increases the optimal absorber thickness, as thinner cells are more sensitive to surface recombination.

As pointed out previously, efficiency as a function of the absorber thickness exhibits a wide maximum. For this reason, in Fig.~\ref{fig:fig9}(b) we include tolerance intervals for each material quality. The intervals are indicated by the colour regions, and show thickness range in which efficiency changes by no more than 1\% relative. For example, if we consider Auger-limited cells (the black lines), and assume front SRV equal to $100\,\mathrm{cm/s}$, we can see that the efficiency changes by no more than 1\% relative in the thickness range from 80 to $500\,\mu\mathrm{m}$. We note that the calculations are performed up to $500\,\mu\mathrm{m}$ (which we consider a practical limit for fabrication).
 
The results presented above show that the optimal absorber thickness is determined by an interplay between bulk and surface losses. Nevertheless, for a reasonably high bulk and surface quality (consistent with present-day fabrication techniques), an optimal thickness is always above $100\,\mu\mathrm{m}$. Therefore, the conclusion that decreasing the absorber thickness below a few tens of micrometers is counter-productive holds for a wide range of material parameters. However, we emphasize that reducing parasitic losses is required in order to increase the energy conversion efficiency above 26\%.


\section{Conclusions}
\label{sec:conclusions}

In conclusion, we presented an electro-optical framework to calculate the limiting efficiency of c-Si solar cells. In our approach, we take the analytical photogeneration rate (assuming full or partial light trapping) and numerically solve the drift-diffusion equations to obtain the cell performance. This gives a realistic description of carrier dynamics in the device and allows easily introducing intrinsic and extrinsic loss mechanisms. Comparison with present-day HIT structures suggests that silicon solar cells are limited by (in decreasing order of significance) $J_\mathrm{sc}$, FF, and $V_\mathrm{oc}$. Thus, improving $J_\mathrm{sc}$ (via light trapping and reducing parasitic losses) and FF (by improving the quality of contacts) is a key to further boost the performance beyond the current efficiency record.

In the first part of this contribution, we calculated the efficiency limits of c-Si solar cells. Therefore, we focused on the cells limited by intrinsic Auger recombination. We have demonstrated that the limiting efficiency as a function of the absorber thickness exhibits a wide maximum: $40\,\mu\mathrm{m}$ thick cell can be nearly as efficient as the solar cell with the optimal absorber thickness (around $170\,\mu\mathrm{m}$). It is therefore more practical to consider the \textit{optimal thickness range}, rather than a single optimal thickness. In this regard, we argued that decreasing the thickness below around $40\,\mu\mathrm{m}$ is counter-productive, as it significantly reduces the maximum achievable efficiency.

When extrinsic SRH recombination is considered, we notice that thicker cells are more sensitive to bulk losses, and therefore the conditions to reach the limiting efficiency for thinner cells are relaxed. Yet, including surface recombination shows the opposite trend: thicker cells are less sensitive to surface recombination, as the surface-to-volume ratio is reduced. Nevertheless, for a reasonably high bulk and surface quality, it remains true that decreasing the thickness below a few tens of micrometers is counter-productive in terms of the efficiency.


The current efficiency record for single junction silicon solar cells is 25.6\% \cite{hit2014,green2015solar}. Based on the analysis presented in this paper, we believe that the efficiency above 26\% is a reasonable next step, which is within the reach of present-day fabrication techniques. Yet, this requires, in the first place, reduction of parasitic losses. Surpassing an efficiency of around 27\% puts severe constraints on surface recombination. Increasing the efficiency above the calculated limit for single junction c-Si solar cells requires novel technologies, like silicon-perovskite tandem structures \cite{malinkiewicz2014perovskite,white2014tandem,bailie2015semi,loper2014organic}.




\appendix

\section{Accuracy of the idealized diode equation}

The efficiency limits of silicon solar cells are usually calculated using idealized diode equation \cite{richter2013reassessment}:

\begin{equation}
J(V) = J_\mathrm{L} - qWR,
\label{eq:diode_eq}
\end{equation}

\noindent where $J_\mathrm{L}$ is the photogenerated current density, $W$ is the cell thickness, and $R$ is the total recombination rate. In this appendix, we elaborate on the difference between the results obtained using Eq.~(\ref{eq:diode_eq}) and numerical solution of drift-diffusion equations by means of finite-element approximation. This allows us to investigate the accuracy of the results obtained using Eq.~(\ref{eq:diode_eq}) in a wide absorber thickness range.

\begin{figure}[!t]
\includegraphics[width=1.0\textwidth]{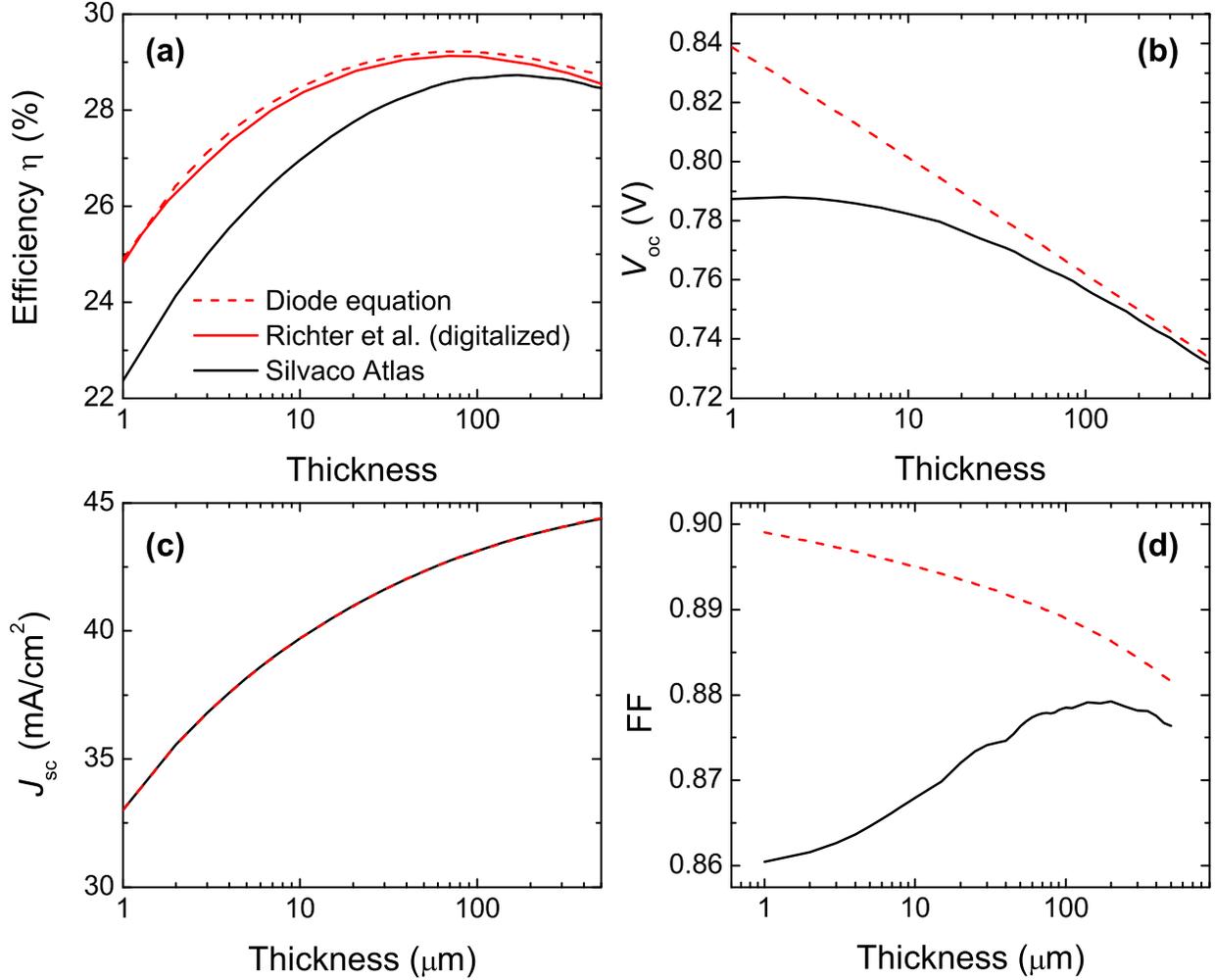}
\caption{\label{fig:fig_appendix} (a) Efficiency, (b) open-circuit voltage $V_\mathrm{oc}$, (c) short-circuit current $J_\mathrm{sc}$, and (d) fill factor FF as a function of absorber thickness. The red dashed lines refer to the results obtained using the idealized diode equation, whereas the black lines refer to the denote the results obtained by solving the drift-diffusion equations by means of finite-element approximation. In (a) we show the comparison with the results digitalized from Ref.~\cite{richter2013reassessment}.}
\end{figure}

In Fig.~\ref{fig:fig_appendix} we show (a) efficiency, (b) open-circuit voltage $V_\mathrm{oc}$, (c) short-circuit current $J_\mathrm{sc}$, and (d) fill factor FF as a function of the absorber thickness. The red dashed lines refer to the results obtained using the idealized diode equation, whereas the black lines denote the results obtained by solving the drift-diffusion equations by means of finite-element approximation. The material parameters and recombination models are the same in both approaches: we include Auger recombination according to Richter \textit{et al.} \cite{richter2012improved} and BGN according to the model by Schenk \cite{schenk1998finite}.

By performing the calculations using Eq.~(\ref{eq:diode_eq}), we aim at reproducing the results reported in Ref.~\cite{richter2013reassessment}. In Fig.~\ref{fig:fig_appendix}(a) we compare our calculations of efficiency with the results presented in the reference work. The results obtained using Eq.~(\ref{eq:diode_eq}) are very close to the results digitalized from Ref.~\cite{richter2013reassessment}. We attribute small discrepancies to the fact that in our calculations we neglect free-carrier absorption and photon recycling: as discussed above, these two effects are likely to compensate each other, and indeed their impact on the cell performance is negligible. For this reason, we have also neglected these effects in the calculations presented in the main part of this paper. Finally, the calculations presented in this work are performed at temperature $T=300\,\mathrm{K}$, whereas the results in the reference work are performed for $T=25^{\circ}\mathrm{C}$.

The discrepancy between the results obtained using Eq.~(\ref{eq:diode_eq}) and full numerical simulations increases with decreasing absorber thickness, as demonstrated in Fig.~\ref{fig:fig_appendix}. In the limiting case of very thick cells, both approaches give essentially the same results. Yet, the difference in efficiency for $1\,\mu\mathrm{m}$ thick cell is close to 2\% (absolute value), i.e., the ideal diode treatment gives higher efficiency. Since $J_\mathrm{sc}$ calculated using both approaches is the same \footnote{The fact that $J_\mathrm{sc}$ is the same in both cases also confirms that the mesh used in the finite-element calculations is dense enough to properly reproduce the photogeneration profile.}, we can trace the discrepancy back to the difference in $V_\mathrm{oc}$ and fill factor: $V_\mathrm{oc}$ calculated using Eq.~(\ref{eq:diode_eq}) increases linearly with decreasing thickness, whereas  $V_\mathrm{oc}$ obtained from the full numerical simulations tends to saturate. Moreover, FF calculated using Eq.~(\ref{eq:diode_eq}) increases with decreasing absorber thickness, whereas FF obtained from the full numerical simulations shows the opposite trend (small fluctuations may be due to numerical inaccuracy).

We can therefore conclude that the results obtained using the idealized diode equation become less accurate with decreasing absorber thickness, i.e., in this approach the efficiency for thin cells is overestimated. We attribute this inaccuracy to the assumptions regarding the treatment of the space-charge region (SCR) in the idealized diode formalism: the idealized diode equation can be derived from the drift-diffusion equations assuming that there is no carrier generation nor recombination in SCR \cite{ashcroft_mermin}. It means that the currents in SCR are constant. With the assumed doping concentrations, the width of SCR is around $350\,\mathrm{nm}$. Therefore, if the cells are thick, SCR is only a small part of the whole cell. Yet, in the case of thin cells, SCR is a significant part of the whole cell, and the assumption that there is no carrier generation nor recombination in SCR seriously disturbs the current distribution in the cell.

\section*{Acknowledgements}

This work was supported by the EU through Marie Curie Action FP7-PEOPLE-2010-ITN Project No. 264687 ''PROPHET''. The authors are grateful to Angelo Bozzola for carefully reading the manuscript and Marco Liscidini for many fruitful discussions.


\providecommand{\noopsort}[1]{}\providecommand{\singleletter}[1]{#1}%

\end{document}